\documentclass[notoc]{JHEP3}

\usepackage{graphicx}
\usepackage{bm}
\input mathrsfs.sty

\def\Tr{\mbox{Tr}\,}

\def\D{{\mathscr D}}
\def\half{\frac{1}{2}}
\def\d{\partial}
\def\slash#1{\, /\kern-0.6em{#1}}

\title{Flux dualization in broken SU(2)}

\author{Chandrasekhar Chatterjee, Amitabha Lahiri\\
\email{chandra@bose.res.in}, \email{amitabha@bose.res.in}\\ 
Department of Theoretical Sciences\\ 
S.~N.~Bose National Centre for Basic Sciences\\ 
Block JD, Sector III, Salt Lake, Kolkata 700 098, W.B. India. }

\abstract{An SU(2) gauge theory is broken to U(1) by an adjoint
  scalar to produce magnetic monopoles. At a lower scale, this U(1)
  is further broken by a fundamental scalar to produce tubes of
  magnetic flux. We dualize the resulting theory to write an
  effective theory in terms of the macroscopic string variables.
  The monopoles are attached to the ends of the strings, and the
  flux is confined in the tubes. }

\begin{document}

\section{\label{sec:level1}Introduction:}

It is widely believed that color confinement in the strong coupling
regime should be a phenomenon dual to monopole confinement in a
color superconductor at weak coupling. In this picture, the QCD
vacuum behaves like a dual superconductor, created by condensation
of magnetic monopoles, in which confinement is analogous to a dual
Meissner effect. Quarks are then bound to the ends of a flux
string~\cite{ Mandelstam:1974pi, Nambu:1975ba, Nambu:1974zg}
analogous to the Abrikosov-Nielsen-Olesen vortex string of Abelian
gauge theory~\cite{Abrikosov:1956sx, Nielsen:1973cs}.

A construction of flux strings in the Weinberg-Salam theory was
suggested by Nambu~\cite{Nambu:1977ag}, in which a pair of magnetic
monopoles are bound by a flux string of Z condensate. The magnetic
monopoles are introduced by hand. If we demand that the magnetic
monopoles should appear from the underlying gauge theory, we need
an additional adjoint scalar field. Such a construction of flux
string, involving two adjoint scalar fields in an SU(2) gauge
theory, has been discussed in~\cite{Nielsen:1973cs, deVega:1976rt}.
Recently there has been a resurgence of interest in such
constructions~\cite{Auzzi:2003fs, Hanany:2004ea, Shifman:2002yi,
  'tHooft:1999au}. We have previously shown explicitly
that~\cite{Chatterjee:2009pi} an SU(2) gauge theory broken by two
adjoint scalar fields at different energy scales has configurations
of magnetic monopoles bound by flux strings.

In this paper we consider an SU(2) gauge theory coupled to an
adjoint scalar field as well as a fundamental scalar field. The two
fields break the symmetry at two scales. At the higher scale the
adjoint scalar breaks the symmetry down to U(1) and produces 't
Hooft-Polyakov magnetic monopoles~\cite{'tHooft:1974qc,
  Polyakov:1974ek, Prasad:1975kr}. The fundamental scalar breaks
the remaining U(1) symmetry at a lower scale and produces a flux
string.

Our starting point is the Lagrangian
\begin{eqnarray}
  \label{lagrangian_1} 
  {\it{ L}} = - {\frac
    12}\Tr\left(G_{\mu\nu}G^{\mu\nu}\right) + \Tr\left(D_\mu \phi
    D^\mu     \phi \right) + \half(D_\mu \psi^{\dagger})(D^\mu
  \psi) + V(\phi,   \psi).  
\end{eqnarray} 
Here $\phi$ is in the adjoint representation of $SU(2), \phi =
\phi^i\tau^i$ with real $\phi^i\,$ and $\psi$ is a fundamental
doublet of $SU(2)$, with $V(\phi, \psi)$ some interaction potential
for the scalars. The $SU(2)$ generators $\tau^i$ satisfy
$\Tr(\tau^i\tau^j) = \half\delta^{ij}$. The covariant derivative
$D_\mu$ and the Yang-Mills field strength tensor $G_{\mu\nu}$ are
defined as
\begin{eqnarray}
  \left(D_\mu\phi\right)^i &=& \partial_\mu \phi^i + g
  \epsilon^{ijk}A_{\mu}^j\phi^k\,, \hfil\\ 
  G_{\mu\nu}^i &=& \partial_\mu {A^i}_{\nu}- \partial_\nu {A^i}_{\mu}
  + g \epsilon^{ijk}{A^j}_{\mu}{A^k}_{\nu}\,,\hfil\\
  (D_\mu \psi)_\alpha &=& \d_\mu \psi_\alpha -
  igA_\mu^i\tau^i_{\alpha\beta} \psi_\beta\,. 
\end{eqnarray}
We will sometimes employ vector notation, in which
\begin{eqnarray}
D_\mu\vec\phi &=& \partial_\mu\vec\phi + g\vec A
\times \vec\phi\,, \\
D_\mu\psi &=& \partial_\mu \psi - ig A_\mu\psi\,,\\
\vec G_{\mu\nu}&=& \partial_\mu\vec A_\nu -
\partial_\nu \vec A_\mu + g \vec A_\mu \times\vec A_\nu\,,\, {\rm 
  etc.} 
\end{eqnarray}
Obviously, $\vec\phi$ and $\phi$ represent the same object. The
simplest form of the potential $V(\phi, \psi)$ that will serve
our purpose is,
\begin{eqnarray}
\label{potential}
V(\phi, \psi) = - {\frac{\lambda_1} 4}(|\phi|^2 - v_1^2)^2  
-  {\frac{\lambda_2} 4}(\psi^\dagger\psi - v_2^2)^2  -
V_{mix}(\phi,\psi). 
\end{eqnarray}
Here $v_1\,,v_2$ are the parameters of dimension of mass and
$\lambda_1\,, \lambda_2$ are dimensionless coupling constants. The
last term $V_{mix}(\phi,\psi)$ includes all mixing terms in the
potential, which involve products of the two scalar fields in some
way.  We will take $V_{mix}(\phi\,,\psi) = 0$ for now, so $v_1$ and
$v_2$ are the local minima of the potential, and we will refer to
them as the vacuum expectation values of $\phi$ and $\psi$.

The adjoint scalar $\phi$ acquires a vacuum expectation value (vev)
$\vec v_1$ which is a vector in internal space, and breaks the
symmetry group down to U(1). The 't Hooft-Polyakov monopoles are
associated with this breaking.  The other scalar field $\psi$ also
has a non-vanishing vev $v_2$ which is a vector in the fundamental
representation. This vector can be associated uniquely with a
vector in the adjoint space which is free to wind around $\vec
v_1$.  A circle in space is mapped to this winding, giving rise to
the vortex string.  We then dualize the fields as
in~\cite{Davis:1988rw, Mathur:1991ip, Lee:1993ty, Akhmedov:1995mw,
  Chatterjee:2006iq} to write the action in terms of string
variables.

The idea of two-scale symmetry breaking in SU(2), the first to
produce monopoles and the second to produce strings, has appeared
earlier~\cite{Hindmarsh:1985xc}. Later this idea was used in a
supersymmetric setting in~\cite{Kneipp:2003ue, Auzzi:2003em,
  Eto:2006dx}, where the idea of flux matching, following
Nambu~\cite{Nambu:1977ag} was also included. The model we discuss
in this paper, with one adjoint and one fundamental scalar, has
been considered previously in~\cite{Shifman:2002yi}.  Here we
construct the flux strings explicitly in non-supersymmetric SU(2)
theory with 't Hooft-Polyakov monopoles of the same theory attached
to the ends. The internal direction of symmetry breaking is left
arbitrary, so that the magnetic flux may be chosen to be along any
direction in the internal space. We also dualize the variables to
write the effective theory of macroscopic string variables coupled
to an antisymmetric tensor, and thus show explicitly that the flux
at each end of the string is saturated by the magnetic monopoles,
indicating confinement of magnetic flux.

\section{Magnetic monopoles}
We assume that $v_1\,,$ the vacuum expectation value of $\phi\,,$
is large compared to the energy scale we are interested in. Below
the scale $v_1\,,$ we find the $\phi$ vacuum, defined by the
equations 
\begin{eqnarray}
\label{1sthiggsvacuum}
D_\mu{\vec\phi} &=& 0\,,\\
|\phi|^2 &=& v_1^2.\nonumber
\end{eqnarray}
Below $v_1\,,$ the original SU(2) symmetry of the theory is broken
down to U(1).  At low energies the theory is essentially Abelian,
with the component of $A$ along $\phi$ remaining massless.  We can
now write the gauge field below the scale $v_1$ as
\begin{eqnarray}
\label{A_phivac}
\vec A_\mu = B_{\mu}\hat{\phi}_1 - {\frac 1g}
\hat{\phi}_1\times\d_{\mu} \hat{\phi}_1\,,
\end{eqnarray}
where $B_{\mu} = \vec A_{\mu}\cdot \hat{\phi}_1$ and $\hat\phi_1 =
\vec\phi_1/v_1$~\cite{Corrigan:1975hd}. In this vacuum, until we
include the second symmetry breaking, $B_\mu$ is a massless
mode. The other two components of $A\,,$ which we call $A^\pm\,,$
and the modulus of the scalar field $\phi$ acquire masses,
\begin{eqnarray}
M_{A^\pm} = gv_1, \qquad M_{|\phi|} = \sqrt \lambda_1 v_1.
\end{eqnarray}
Well below $v_1$ the modes $A^\pm$ are not excited, so they will
not appear in the low energy theory. The second term on the right
hand side of Eq.~(\ref{A_phivac}) corresponds to the gauge field
for SU(2) magnetic monopoles~\cite{'tHooft:1999au}.

A straightforward calculation shows that,
\begin{eqnarray}
\Tr (G_{\mu\nu}G^{\mu\nu}) &=& {\frac 12}F_{\mu\nu}F^{\mu\nu},
\end{eqnarray}
where
\begin{eqnarray}
\label{F}
F_{\mu\nu} &=& \d_{[\mu}B_{\nu]} - 
{\frac 1g}\hat{\phi}_1\cdot\d_\mu \hat{\phi}_1\times
\d_\nu\hat{\phi}_1 \, \equiv \partial_{[\mu} B_{\nu]} + M_{\mu\nu}.
\end{eqnarray}
Then the Lagrangian can be written in the $\phi$-vacuum as 
\begin{eqnarray}
L = -  {\frac 14}F_{\mu\nu}F^{\mu\nu} 
+ (D_\mu \psi^{\dagger})(D^\mu \psi)
 -  {\frac{\lambda_2} 4}(\psi^\dagger \psi - v_2^2)^2 .
\end{eqnarray}

The second term of Eq.~(\ref{F}) is the `monopole term'.  In a
configuration where the scalar field at spatial infinity goes as
$\phi_1^i \to v_1\displaystyle{\frac {r^i}{r}}$, the $(ij)^{th}$
component of the second term of Eq.~(\ref{F}) becomes
$-\displaystyle{{\frac{\epsilon_{ijk} r^k}{gr^3}}},$ which we can
easily identify as the field of a magnetic monopole.  The flux for
this monopole field is ${4\pi}\over g $.  On the other hand, a
monopole with magnetic charge $Q_m$ produces a flux of $4\pi Q_m,$
and thus we find the quantization condition for unit charge, $Q_m g
= 1.$

The scalar field $ \phi $ can be written as $\phi(x) =
|\phi(x)|\hat\phi(x)$, where $\hat\phi$ contains two independent
fields (and $x\equiv \vec x$). So under a gauge transformation
$\hat\phi$ has a trajectory on $ S^2 $. Since $\phi$ is in the
adjoint of SU(2), we can always write $\phi$ as
\begin{equation}
\phi(x) = |\phi(x) | g(x)\tau ^3 g^{-1}(x) = |\phi(x)| \hat\phi(x) 
\,,
\end{equation}
with $g(x) \in $ SU(2). Then for a given $\phi(x)\,,$ we can
locally decompose $g(x)$ as $g(x) = h(x)U(x)\,,$ with $h(x) = \exp
(- i\xi (x)\hat\phi(x))\,,$ and we can write
\begin{eqnarray}
\phi(x) = |\phi(x)| U(\varphi(x), \theta(x))\tau ^3
U^\dagger(\varphi(x), \theta(x)), 
\label{hU}
\end{eqnarray}
Here $\xi(x), \varphi(x), \theta(x)$ are angles on $S^3$=
SU(2). The matrix $U$ rotates $\hat\phi(x)$ in the internal space,
and is an element of SU(2)/U(1), where the U(1) is the one
generated by $h\,.$ If $|\phi|$ is zero at the origin and $ |\phi|$
goes smoothly to its vacuum value $v_1 $ on the sphere at infinity,
the field $ \phi $ defines a map from space to the vacuum manifold
such that second homotopy group of the mapping is ${\mathbb Z}$.
Equating $ \phi $ with the unit radius vector of a sphere we can
solve for $U(\theta(x),\varphi (x))$,
\begin{eqnarray}
\label{Umonopole1}
U = \left(\begin{tabular}{cc}
$\cos{\theta\over 2} $ & $-\sin{\theta\over 2}e^{-i \varphi}$ \\
$\sin{\theta\over 2}e^{i \varphi}$ & $\cos{\theta\over 2}$\\
\end{tabular}\right)\, . 
\end{eqnarray}

An 't~Hooft-Polyakov monopole (in the point approximation, or as
seen from infinity) at the origin is described by
\begin{eqnarray}
\label{Umonopole2}
 U = \cos{\theta\over 2}\left(\begin{tabular}{cc}
$e^{i \varphi}$ & 0 \\
0 & $e^{-i \varphi}$\\ 
\end{tabular}\right) + \sin{\theta\over 2}
\left( \begin{tabular}{cc}  
$0\quad$ & $i$\\
$i\quad$ & $0$\\ 
\end{tabular}\right)\,, 
\end{eqnarray}
where $0\le\theta(\vec x)\le \pi$ and $0\le\varphi(\vec x)\le2\pi$
are two parameters on the group manifold.  This choice of $U(\vec
x)$ is different from that in Eq.~(\ref{Umonopole1}) by a rotation
of the axes.  Both choices lead to the field configuration
\begin{eqnarray}
\vec\phi &=&  v_1{\frac {r^i}{r}}\tau_i.
\end{eqnarray}
For this case, $Q_m g = 1\,,$ as we mentioned earlier. A monopole
of charge $n/g$ is obtained by making the replacement $\varphi \to
n\varphi $ in Eq.s~(\ref{Umonopole1}, \ref{Umonopole2}).
%
%
The integer $n$ labels the homotopy class, $\pi_2(SU(2)/U(1))
\sim \pi_2(S^2) \sim Z\,,$ of the scalar field configuration.
Other choices of $U(\vec x)$ can give other configurations. For
example, a monopole-anti-monopole
configuration~\cite{Bais:1976fr} is given by the
choice
\begin{eqnarray}
\label{M-anti-M}
U = \sin{({\theta_1 - \theta_2})\over 2}\left(\begin{tabular}{cc} 
0 &   $- e^{ -i \varphi}$\\
$e^{i \varphi}$     & 0\\ 
\end{tabular}\right) + \cos{({\theta_1 - \theta_2})\over 2}
\left( \begin{tabular}{cc}  
 $1\quad$ & $0$\\
 $0\quad$ & $1$\\ 
 \end{tabular}\right) . 
\end{eqnarray}
For our purposes, we will need to consider a $\phi_1$-vacuum
configuration with $U(\vec x) \in SU(2)$ corresponding to a
monopole-anti-monopole pair separated from each other by a distance
$> 1/v_1.$ Then the total magnetic charge vanishes, but each
monopole (or anti-monopole) can be treated as a point particle.

\section{Flux tubes}
We started with a theory with SU(2) symmetry and a pair of scalars
$\phi,\psi\,.$ The non zero vacuum expectation value $v_1$ of the
field $\phi$ breaks the symmetry to U(1), so that below $v_1$ we
have an effective Abelian theory with magnetic monopoles.  The
gauge group SU(2) acts transitively on the vacuum manifold $S^2$,
so the Abelian effective theory is independent of the internal
direction of $\phi$. The remaining symmetry of the theory is the
U(1), the little group of invariance of $\phi$ on the vacuum
manifold.  This is the group of rotations around any point on the
vacuum manifold $S^2$.

There is another scalar field $\psi$ in the theory, a scalar in the
fundamental representation of SU(2). After breaking the original
SU(2) down to the $\phi$-vacuum, the only remaining gauge symmetry
of the SU(2) doublet $\psi$ is a transformation by the little group
U(1). We will find flux tubes when this U(1) symmetry is
spontaneously broken down to nothing. The elements of this U(1) are
$h(x) = \exp[i\xi (x)\hat\phi(x)]\,,$ rotations by an angle
$\xi(x)$ around the direction of $\phi(x)$ at any point in
space. This U(1) will be broken by the vacuum configuration of
$\psi\,.$  



Let us then define the $\psi$-vacuum by,
\begin{eqnarray}
 \label{2ndhiggs1}
 \psi^{*i}\psi^i = v_2^2\\
 \label{2ndhiggs2}
 D_\mu \psi = 0,
\end{eqnarray}
where $ D_\mu $ is defined using $ A_\mu $ in the $\phi$-vacuum, as
in Eq.~(\ref{A_phivac}).  Multiplying Eq.~(\ref{2ndhiggs2}) by
$\psi^\dagger\hat\phi$ from the left, its adjoint by
$\hat\phi\psi$ from the right, and adding the results, we get
\begin{eqnarray}
0 & =& \psi^\dagger\hat\phi D_\mu\psi  +
(D_\mu\psi^\dagger )  \hat\phi\psi\, \nonumber\\
&= & \d_\mu\left[\psi^\dagger\hat\phi\psi\right]\,,
\end{eqnarray}
from which it follows that 
\begin{eqnarray}
\label{phi}
\psi^\dagger\hat\phi\psi = \mathrm{constant}\,,
\end{eqnarray}
or explicitly in terms of the components,
\begin{eqnarray}
\label{psi1}
\Tr\left[{\psi^\dagger}_i\sigma^\alpha_{ij}\psi_j
  \tau_\alpha\hat\phi\right] = \mathrm{constant}\,.
\end{eqnarray} 
It follows that the components parallel and orthogonal to $\phi$
are both constants.  Then we can decompose
\begin{eqnarray}
\label{psi2}
{\psi^\dagger}_i\sigma^\alpha_{ij}\psi_j \tau_\alpha =
v^2_2\cos\theta_c \hat\phi +  v^2_2 \sin\theta_c \hat \kappa\,,
\end{eqnarray}
where $\hat\kappa$ is a vector in the adjoint, orthogonal to
$\hat\phi\,.$ We can always write $\hat\kappa$ as
\begin{eqnarray}
\label{h}
\hat\kappa = hU\tau^2U^\dagger h^\dagger\,,
\end{eqnarray}
where $h$ and $U$ are as defined before and in Eq.~(\ref{hU}).

Using the identity $ {\sigma^\alpha}_{ij} {\sigma^\alpha}_{kl} =
\delta_{il}\delta_{kj} - \half \delta_{ij}\delta_{kl}$, we find
that $\psi$ is a eigenvector of the expression on the left hand
side of Eq.~(\ref{psi2}). Then writing the right hand side of that
equation in terms of $h$ and $U$, we find that $\psi$ can be
written as
\begin{eqnarray}
\label{psi1}
\psi = v_2hU \left(\begin{tabular}{l}
                                 $\rho_1$\\
                                  $\rho_2$\\
                         \end{tabular}\right)\,,
\end{eqnarray}
%
where $\rho_1$ and $\rho_2$ are constants.  Keeping $U$ fixed, we
vary $\xi$ and find the periodicity
\begin{eqnarray}
\psi(\xi) = \psi(\xi + 4\pi)\,.
\end{eqnarray}
This $\xi$ is the angle parameter of the residual $U(1)$ gauge
symmetry and in the presence of a string solution, this $\xi$ is
mapped a circle around the string. In order to make $\psi$ single
valued around the string, we need $\xi = 2 \chi$, where $ \chi $ is
the angular coordinate for a loop around the string. Next let us
calculate the Lagrangian of the scalar field $\psi$.  We have
\begin{eqnarray}
  D_\mu\psi &=& \d_\mu\psi -igA_\mu\psi\\
  &=& \d_\mu (hU\rho) -ig\left[B_\mu \hat\phi +
    ig\left[\hat\phi,
      \d_\mu\hat\phi\right]\right]hU\rho\\ 
  &=& \d_\mu (Uh_0\rho) -ig\left[B_\mu \hat\phi +
    ig\left[\hat\phi, 
      \d_\mu\hat\phi\right]\right]Uh_0\rho\\ 
  &=& - iUh_0\tau^3\rho \left[2\d_\mu\chi  + g\left(B_\mu
      +N_\mu\right)\right]\,,
\end{eqnarray}
where $h_0 = e^{- i2\chi\tau_3}\,,\, \rho^i\rho^i = {v_2}^2\,,$ and
we have used the identity $ U^\dagger h U = \exp  (- 2i \chi\tau^3)\,.$
We have also introduced the Abelian `monopole field'
\begin{eqnarray}
 N_\mu &=&  2iQ_m \Tr\left[ \d_\mu U U^{\dagger}\hat
   \phi\right]\,,\\ 
\d_{[\mu} N_{\nu]} &=&   Q_mM_{\mu\nu} +
2iQ_m\Tr[(\d_{[\mu}\d_{\nu]}U) 
U^{\dagger}\hat\phi]\,.
\end{eqnarray}
The first term reproduces the magentic field of the monopole
configuration, while the second term is a gauge dependent line
singularity, the Dirac string. This singular string is a red
herring, and we are going to ignore it because it is an artifact of
our construction. We have used a $U(\vec x)$ which is appropriate
for a point monopole. If we look at the system from far away, the
monopoles will look like point objects and it would seem that we
should find Dirac strings attached to each of them. However, we
know that the 't~Hooft-Polyakov monopoles are actually not point
objects, and their near magnetic field is not describable by an
Abelian four-potential $N_\mu,$ so if we could do our calculations
without the far-field approximation, we would not find a Dirac
string. Further, as was pointed out in~\cite{Chatterjee:2009pi},
the actual flux tube occurs along the line of vanishing $\psi\,,$
and it is always possible to choose a $U(\vec x)$ appropriate for
the monopole configuration such that the Dirac string lies along
the zeroes of $\psi$. Since $|\psi|^2$ always multiplies the term
containing $N_\mu$ in the action, the effect of the Dirac string
can always be ignored.

With these definitions we can
calculate
\begin{eqnarray}
L &=& -{\frac 14}F^{\mu\nu}F_{\mu\nu}   
+ {\frac {{v_2}^2}2}\left(\d_\mu\chi + e\left(B_\mu +
    N_\mu\right)\right)^2
\label{fundareps}
\end{eqnarray}
Here, we have defined electric charge $e = \frac g2$ and written
the magnetic charge as $Q_m = \frac 1{2e}$ .  


\section{Dualization}
Let us now dualize the low energy effective action in order to
express the theory in terms of the macroscopic string variables.
The partition function $Z$ is simply the functional integral
\begin{eqnarray}
Z &=& \int \D B_\mu\D\chi\exp i
\int d^4 x \left[- {\frac 14}F_{\mu\nu}F^{\mu\nu} + 
    {\frac {v_2^2}2}\left(eB_\mu +\d_\mu\chi +
      eN_\mu\right)^2\right]\,.
\label{flux.Higgs}
\end{eqnarray}
In the presence of flux tubes we can decompose the angle $\chi$
into a part $\chi^s$ which measures flux in the tube and a part
$\chi^r$ describing single valued fluctuations around this
configuration, $\chi = \chi^r + \chi^s\,.$ Then if $ \chi $ winds
around the tube $n$ times, we can define
\begin{eqnarray}
\epsilon^{\mu\nu\rho\lambda}\partial_{\rho}
\partial_{\lambda}\chi^s  = 
2\pi n\int_{\Sigma}d\sigma^{\mu\nu}(x(\xi))\,\delta^4(x-x(\xi))  
\equiv  \Sigma^{\mu\nu}\,,
\label{def.sigma}
\end{eqnarray}
where $\xi = (\xi^1, \xi^2)$ are the coordinates on the world-sheet
and $d\sigma^{\mu\nu}(x(\xi)) = \epsilon^{ab}\partial_a
x^\mu \partial_b x^\nu\,.$ The vorticity quantum is $2\pi$ in the
units we are using and $n$ is the winding
number~\cite{Marino:2006mk}.

The integration over $\chi$ has now become integrations over both
$\chi^r$ and $\chi^s\,$.  However $\chi^r$ is a single-valued
field, so it can be absorbed into the gauge field $ B_\mu $ by a
redefinition, or gauge transformation, $B_\mu \to B_\mu
+ \partial_\mu\chi^r$. We can linearize the action by
introducing auxiliary fields $C_\mu, B_{\mu\nu}$ and $A^{m}_\mu$,
\begin{eqnarray}
Z &=& \int \D B_\mu \D C_\mu \D \chi_s \D B_{\mu\nu} \D
A^m_{\mu}\nonumber \\
&& \exp i \int d^4 x 
\left[ -\frac{1}{4} G^{\mu\nu}G_{\mu\nu} + \frac{1}{4}
  \epsilon^{\mu\nu\rho\lambda} 
G_{\mu\nu}F_{\rho\lambda} - \frac{1}{2v_2^2}C_\mu^2 - C^\mu (eB_\mu
+ eN_\mu + \d_\mu \chi_s) \right]\,,\nonumber \\
\end{eqnarray}
where we have written $G_{\mu\nu} = \d_\mu A^m_\nu - \d_\nu A^m_\mu 
+ ev_2 B_{\mu\nu}.$ and $F_{\mu\nu}= \partial_\mu B_\nu
- \partial_\nu B_\mu + M_{\mu\nu}\,$.  Now we can integrate over
$B_\mu$ easily,
\begin{eqnarray}
Z = \int \D C_\mu \D \chi_s \D B_{\mu\nu} \D A^m_{\mu} 
\delta\left( C^\mu - \frac{v_2}{2}
\epsilon^{\mu\nu\rho\lambda}\d_\nu B_{\rho\lambda}\right)
\exp i \int d^4 x \qquad\qquad \qquad\nonumber\\
\left[ -\frac{1}{4} G^{\mu\nu}G_{\mu\nu} + \frac{ev_2}{4}
  \epsilon^{\mu\nu\rho\lambda} 
B_{\mu\nu}M_{\rho\lambda} - A^\mu j_\mu - \frac{1}{2v_2^2}C_\mu^2
 - C^\mu (eB_\mu + eN_\mu + \d_\mu \chi_s) \right]\,.
\end{eqnarray}
Here $j_m^{\mu} = - {\frac 1{2}} \epsilon^{\mu\nu\rho\lambda}\d_\nu
M_{\rho\lambda}$ is the magnetic monopole current.  Integrating
over $ C_\mu $ we get
\begin{eqnarray}
Z = \int \D \chi_s \D B_{\mu\nu} \D A^m_{\mu} 
\exp i \int d^4 x 
\left[ -\frac{1}{4} G^{\mu\nu}G_{\mu\nu} + \frac{1}{12}
  H^{\mu\nu\rho}H_{\mu\nu\rho} -  
\frac{v_2}{2} \Sigma_{\mu\nu}B^{\mu\nu} - A^\mu j_\mu \right],
\end{eqnarray}
where we have written  defined
$H_{\mu\nu\rho} = \partial_\mu B_{\nu\rho} + \partial_\nu
B_{\rho\mu} + \partial_\rho B_{\mu\nu}\,,$ used
Eq.~(\ref{def.sigma}) and also written $M_{\mu\nu} = (\d_\mu N_\nu -
\d_\nu N_\mu)\,.$

We can also replace the integration over $\D\chi^s$ by an
integration over $\D x_\mu(\xi)$, representing a sum over all the
flux tube world sheet where $x_{\mu}(\xi)$ parametrizes the surface
of singularities of $ \chi $. The Jacobian for this change of
variables gives the action for the string on the background space
time~\cite{Akhmedov:1995mw, Orland:1994qt}. The string has a
dynamics given by the Nambu-Goto action, plus higher order
operators~\cite{Polchinski:1991ax}, which can be obtained from the
Jacobian. We will ignore the Jacobian below, but of course it is
necessary to include it if we want to study the dynamics of the
flux tube.
\begin{eqnarray}
  Z = \int \D x_\mu(\xi) \D B_{\mu\nu} \D A^m_{\mu} 
  \exp i \int d^4 x 
  \left[ -\frac{1}{4} G^{\mu\nu}G_{\mu\nu} + \frac{1}{12}
    H^{\mu\nu\rho}H_{\mu\nu\rho} -  
    \frac{v_2}{2} \Sigma_{\mu\nu}B^{\mu\nu} - A^\mu j_\mu \right],
\label{flux.functional}
\end{eqnarray}
The equations of motion for the field $B_{\mu\nu}$ and $A^{\mu}$
can be calculated from this to be
\begin{eqnarray}
\label{flux.Beom}
\partial_\lambda H^{\lambda\mu\nu} &=& -m \, G^{\mu\nu} -
\frac{m}{e} \,\Sigma^{\mu\nu} \,,\\
\d_\mu G^{\mu\nu} &=& j_m^\mu
\label{flux.Aeom} 
\end{eqnarray}
where $G_{\mu\nu}= ev_2  B_{\mu\nu} + \partial_{\mu}A^m_{\nu} -
\partial_{\nu}A^m_{\mu}\,,$ and $m = e v_2$. Combining
Eq.~(\ref{flux.Aeom}) and Eq.~(\ref{flux.Beom}) we find that
\begin{equation}
\frac 1e  \partial_\mu \Sigma^{\mu\nu}(x) + j_m^\mu(x) = 0\,.
\label{mono.coneq}
\end{equation}
It follows rather obviously that a vanishing magnetic monopole
current implies $\partial_\mu \Sigma^{\mu\nu}(x) = 0\,,$ or in
other words if there is no monopole in the system, the flux tubes
will be closed. 

The magnetic flux through the tube is
$\displaystyle{\frac{2n\pi}e}\,,$ while the total magnetic flux of
the monopole is $\displaystyle{\frac{4m\pi}g}\,,$ where $n, m$ are
integers.  Since $\displaystyle{e Q_m = \half} $, it follows that
we can have a string that confine a monopole and anti-monopole pair
for every integer $ n $. Although this string configuration could
be broken by creating a monopole-anti-monopole pair, there is a
hierarchy of energy scales $v_1\gg v_2\,,$ which are respectively
proportional to the mass of the monopole and the energy scale of
the string. So this hierarchy can be expected to prevent string
breakage by pair creation.

The conservation law of Eq.~(\ref{mono.coneq})
also follows directly from $ Z $ in
Eq.~(\ref{flux.functional}) by introducing a variable 
$B'_{\mu\nu} = B_{\mu\nu} +
{\frac 1m } (\partial_\mu A^m_\nu - \partial_\nu A^m_\mu)$ and
integrating over the field $ A^m_\mu $. If we do so
we get
\begin{eqnarray}
\label{confinement}
Z =\int \D x_{\mu}(\xi) \D B'_{\mu\nu}&&
\delta\Big[\frac1e\partial_\mu\Sigma^{\mu\nu}(x)  
+  j^\nu_m(x)\Big] \nonumber\\
&&\exp\left[i\int\left\{\frac 1{12} H_{\mu\nu\rho}H^{\mu\nu\rho} 
- \frac 14  m^2 {B'}^2_{\mu\nu}- 
{\frac m{2e}} \Sigma_{\mu\nu}{B'}^{\mu\nu} \right\}\right] \,, 
\end{eqnarray}
with the delta functional showing the conservation law
(\ref{mono.coneq}).  Thus these strings are analogous to the
confining strings in three dimensions~\cite{Polyakov:1996nc}. There
is no $A^m_\mu$, the only gauge field which is present is
$B'_{\mu\nu}$. This $B'_{\mu\nu}$ field mediates the direct
interaction between the confining strings.


The delta functional in Eq.~(\ref{confinement}) enforces that at
every point of space-time, the monopole current cancels the
currents of the end points of flux tube.  So the monopole current
must be non-zero only at the end of the flux tube.
Eq.~(\ref{confinement}) does not carry Abelian gauge field
$A^m_{\mu}$, only a massive second rank tensor gauge field.  All
this confirms the permanent attachment of monopoles at the end of
the flux tube which does not allow gauge flux to escape out of the
flux tubes.  There are important differences between the results
obtained from this construction and that from using two adjoint
scalars.  The mass of the Abelian photon will be zero for the two
adjoint case if the two adjoint vevs are aligned in the same
direction. But this cannot happen for one adjoint and one
fundamental scalar. Also, in this case flux confinement is possible
for all winding numbers of the string.


\end{document}